\documentclass[12pt]{article}

\pdfoutput = 1
\usepackage{titlesec}
\titlelabel{\thetitle.\quad}
\begin{document}
\date{Today}
\title{{\bf{\Large Thermodynamics and emergent universe }}}

\author{
{\bf {\normalsize Saumya Ghosh}$^{a}
$\thanks{sgsgsaumya@gmail.com, sg14ip041@iiserkol.ac.in}},
{\bf {\normalsize Sunandan Gangopadhyay}
$^{a,b}$\thanks{sunandan.gangopadhyay@gmail.com, sunandan@iiserkol.ac.in, sunandan@associates.iucaa.in}}\\
$^{a}$ {\normalsize Indian Institute of Science Education and Research Kolkata}\\
{\normalsize Mohanpur 741246, Nadia, West Bengal, India }\\
$^{b}${\normalsize Visiting Associate in Inter University Centre for Astronomy \& Astrophysics (IUCAA)}\\
{\normalsize Pune 411007, India}\\
}
\date{}

\maketitle

\begin{abstract}
{\noindent We show that in the isentropic scenario the first order thermodynamical particle creation model gives an emergent universe solution even when the chemical potential is non-zero. However there exists no emergent universe scenario in the second order non-equilibrium theory for the particle creation model. We then point out a correspondence between the particle creation model with barotropic equation of state and the equation of state giving rise to an emergent universe without particle creation in spatially flat FRW cosmology.
  }
\end{abstract}
\vskip 1cm



\section{Introduction}
The standard model of cosmology, namely, the big-bang model is known to have an initial singularity. Attempts to avoid this singularity has been a central area of research. Emergent universe models have attracted considerable interest in this context \cite{SGPaper17}-\cite{SGPaper21}. This is a model which in the infinite past ($t \rightarrow -\infty$) is almost static and then proceeds to have an inflationary behaviour. The first approach that led to an emergent universe scenario involved that of a closed universe with only radiation \cite{SGPaper17}. It was shown later that such a model can also emerge from closed universes having a minimally coupled scalar field with a particular form for the self-interacting potential $V(\phi)$ \cite{SGPaper18, SGPaper19}. An interesting way to obtain an emergent universe was demonstrated later in \cite{SGPaper20}. Here, an ad hoc equation of state, non-linear in form was proposed which led to an emergent universe solution. 

Cosmological models with matter creation have also been a prime area of research to understand the origin of cosmological entropy and matter content. Models have been proposed which could explain the creation of matter but were unable to explain the entropy production \cite{SGPaper1}-\cite{SGPaper7}. An important step was taken in \cite{Prigogine02, Prigogine01} to provide an argument for the production of entropy. The formalism augments the field equations of Einstein by introducing a conservation equation for particles. In a recent paper \cite{Sunandan}, it has been shown that one can have an emergent universe scenario in the context of this formalism which allows for an irreversible creation of matter at the cost of the field of gravitation. 

Recently in \cite{Subenoy} it was shown that the first order formalism of non-equilibrium thermodynamics with dissipation due to particle creation mechanism with a barotropic equation of state leads to an emergent universe scenario. An important point to note is that the chemical potential is set to zero in this analysis. It is however not at all clear whether the emergent solution still exists when the chemical potential is non-vanishing. In this paper we carry out a careful thermodynamic analysis to show explicitly that there exists an emergent universe scenario even in the presence of a non-vanishing chemical potential in the first order formalism of non-equilibrium thermodynamics with matter creation.  
Further, it was discussed in \cite{Subenoy} that the second order theory of non-equilibrium thermodynamics with matter creation also admits an emergent universe scenario. We differ from this result in our paper. We have argued that there can be no emergent universe scenario in the second order theory.

Finally, we have shown an interesting duality between the particle creation model with barotropic equation of state and equation of state (non-linear in form) leading to an emergent universe. We have identified the constants appearing in the non-linear equation of state with the adiabatic index and the effective bulk viscous coefficient. 

The paper is organized as follows. In section 2, we have presented the basic formalism of thermodynamics of the creation of matter in the cosmological context. In section 3, we have discussed the first order formalism of non-equilibrium thermodynamics with matter creation. In section 4, we have discussed the second order formalism of non-equilibrium thermodynamics with matter creation. In section 5, we discuss a correspondence between matter creation model for barotropic fluid and equation of state for emergent universe.  We conclude in section 6.


\section{Thermodynamics of matter creation in cosmology}
Let us consider a volume $V$ containing $N$ particles. From the first law of thermodynamics we have for a closed system \cite{Prigogine02, Prigogine01}
\begin{eqnarray}
dQ = dE+pdV 
\label{1}
\end{eqnarray}
where $dQ$ is the heat received by the system, $E$ is the total internal energy of the system and $p$ is the thermodynamic pressure. 
This can be recast as
\begin{eqnarray}
d({\rho}{V})+pdV =dQ
\label{1a}
\end{eqnarray}
with $\rho = E/V$ being the energy density.
For adiabatic closed systems, $dQ = 0$ and this equation takes the form
\begin{eqnarray}
d({\rho}{V})+pdV = 0.
\label{2}
\end{eqnarray}
The standard equations of general relativity read
\begin{eqnarray}
G_{\mu\nu} = {\kappa}T_{\mu\nu}
\label{einstein}
\end{eqnarray} 
where $\kappa=8\pi G$ and we set $\kappa=1$ in the subsequent discussion.
Here $T_{\mu\nu}$ is the stress tensor corresponding to an ideal fluid
\begin{eqnarray}
{T^0}_0 = \rho  ~~;~~    {T^i}_j = p{{\delta}^i}_j ~.
\end{eqnarray}
Einstein's equations when combined with the Bianchi identities 
leads to eq.({\ref{2}}) for homogeneous and isotropic universe with $V$ representing the comoving volume.

\noindent In case of open systems where the particle number is not conserved, eq.({\ref{1a}}) needs to be modified \cite{Prigogine02, Prigogine01}
\begin{eqnarray}
\label{3}
d({\rho}{V})+pdV-\left(\frac{h}{n}\right)d({n}{V}) = dQ
\end{eqnarray}
where $ h = \rho+ p$ is the enthalpy per unit volume and $n=N/V$ is the number density of particles. For adiabatic transformation ($dQ=0$), eq.({\ref{3}}) takes the form
\begin{eqnarray}
\label{4}
d({\rho}{V})+pdV-\left(\frac{h}{n}\right)d({n}{V}) = 0.
\end{eqnarray}
Matter creation corresponds to an additional pressure ${\Pi}$ and therefore the true thermodynamical pressure becomes $(p+\Pi)$. 
This can be seen from eq.({\ref{4}}) as
\begin{eqnarray}
d({\rho}{V}) &=& -\left[p-{\frac{h}{n}}{\frac{d(nV)}{dV}}\right]dV \nonumber\\
&=&-[p+\Pi]dV \equiv -\tilde{p}dV
\label{4a}
\end{eqnarray}
where
\begin{eqnarray}
\label{8}
\Pi = -{\frac{h}{n}}{\frac{d({n}{V})}{dV}}~.
\end{eqnarray}
Eq.(\ref{4a}) is similar in form to that of a closed adiabatic system (\ref{2})
but with the presence of an additional pressure ${\Pi}$ due to matter creation.
Choosing 
\begin{eqnarray}
\label{9}
V = a^{3}{(t)}
\end{eqnarray}
for an isotropic and homogeneous universe leads to 
\begin{eqnarray}
\label{10}
\Pi = -{\frac{{\rho}+p}{3nH}}({\dot{n}+3nH})
\end{eqnarray}
where $a(t)$ is the scale factor and $H=\dot{a}/a$ is the Hubble parameter. 

\noindent The creation of matter for an open thermodynamical system is mathematically expressed as \cite{Prigogine02, Prigogine01}
\begin{eqnarray}
\label{Y}
\dot{n}+3Hn = \psi
\end{eqnarray}
where $\psi$ is the matter creation rate. 
Putting $\psi = n\Gamma$ in the above equation and rearranging, we have
\begin{eqnarray}
\label{Y1}
\frac{\dot{n}}{n}+3\frac{\dot{a}}{a} = \Gamma.
\end{eqnarray}
Using this in eq.({\ref{10}), we get
\begin{eqnarray}
\label{Z}
\Pi = -\frac{\Gamma}{3H}(\rho+p)=-\frac{\psi}{3nH}(\rho+p).
\end{eqnarray}
The above equation implies that the additional pressure $\Pi$ gets determined
by the rate of particle production $\Gamma$.

\noindent The thermodynamical relation (\ref{4a}) can be recast in the form
\begin{eqnarray}
d\rho+(\rho+p+\Pi)\frac{dV}{V} = 0
\label{zz}
\end{eqnarray}
which on using eq.(\ref{9}) leads to 
\begin{eqnarray}
\label{K}
\dot{\rho}+3H(\rho+p+\Pi) = 0.
\end{eqnarray}
Eq.({\ref{Z}}) can also be obtained from the Gibbs equation 
\begin{eqnarray}
Tds = dq = d\left(\frac{\rho}{n}\right)+pd\left(\frac{1}{n}\right)
\label{gibbs}
\end{eqnarray}
where $ds=dS/N$ and $dq=dQ/N$.
The particle number conservation equation (\ref{Y1}) together with 
the energy conservation eq.(\ref{K}) can now be used to obtain the entropy variation from the Gibbs relation (\ref{gibbs}). This reads
\begin{eqnarray}
nT\dot{s} = -3H\Pi-\Gamma(\rho+p)
\label{var}
\end{eqnarray}
with $T$ being the temperature of the fluid. In case of an adiabatic process the entropy per particle $s$ does not change and hence we recover eq.(\ref{Z}).

\noindent Eq.(\ref{K}) is also a consequence of the 
conservation of the stress tensor 
\begin{eqnarray}
\nabla_{\nu}T^{\mu\nu} =0
\label{v1}
\end{eqnarray}
where the stress tensor $T^{\mu\nu}$ is given by 
\begin{eqnarray}
T^{\mu\nu} = (\rho+p+\Pi)u^{\mu}u^{\nu}+(p+\Pi)g^{\mu\nu}
\label{st}
\end{eqnarray}
with $u^{\mu}$ being the particle $4$-velocity.
Note that $\Pi$ is said to be the bulk viscous pressure generating due to particle creation. With this form of the stress tensor, the field equations of Einstein read
\begin{eqnarray}
\label{A}
3H^2 = \rho
\end{eqnarray}
\begin{eqnarray}
\label{B}
-2\dot{H} = \rho+p+\Pi~.
\end{eqnarray}
Substituting the form of the bulk viscous pressure $\Pi$ (eq.\ref{Z}) 
in eq.(\ref{B}) and then using eq.(\ref{A}) along with the barotropic equation of state 
\begin{eqnarray}
\label{E}
p = (\gamma-1)\rho
\end{eqnarray}
with $\gamma$ being the adiabatic index, yields 
\begin{eqnarray}
-2\dot{H} = 3\gamma H^2-\gamma \Gamma H~.
\end{eqnarray}
The above equation can be recast as 
\begin{eqnarray}
\label{C}
\frac{\Gamma}{3H} = 1+\frac{2}{3\gamma}\frac{\dot{H}}{H^2}~.
\end{eqnarray}

\section{First order formalism}
In this section we investigate whether there exists an emergent universe scenario in the first order formalism of non-equilibrium thermodynamics \cite{Eckart}.
The entropy flow vector in this theory is defined as 
\begin{eqnarray}
s^{\mu} = nsu^{\mu}.
\end{eqnarray}
Using the conservation equation for particles (\ref{Y}) and eq.(\ref{Z}) which is obeyed for isentropic transformations, one gets 
\begin{eqnarray}
\nabla_{\mu}{s^{\mu}} = -\frac{\Pi}{T}\left(3H+\frac{\mu n\Gamma}{\Pi}\right)
\label{100}
\end{eqnarray}
where the chemical potential $\mu$ is given by the Euler's relation
\begin{eqnarray}
\label{AA}
\mu = \frac{\rho + p}{n}-T s.
\end{eqnarray}
Now for the second law of thermodynamics to hold ($\nabla_{\mu}{s^{\mu}}{\geq} 0$), one assumes {\cite{Eckart}}
\begin{eqnarray}
\label{Z1}
{\Pi} = -{\zeta}\left({3H}+{\frac{\mu n\Gamma}{\Pi}}\right)
\end{eqnarray}
where ${\zeta}$ is called the coefficient of bulk viscosity and is a constant. Using the above relation in eq.(\ref{100}) gives
\begin{eqnarray}
\nabla_{\mu}{s^{\mu}} = \frac{{\Pi}^2}{T{\zeta}} & {\geq} & 0.
\label{en}
\end{eqnarray}
Eq.({\ref{Z1}}) can be recast as
\begin{eqnarray}
{\Pi}^2+3{\zeta}H{\Pi} = -{\zeta}{\mu}n{\Gamma}.
\label{en11}
\end{eqnarray}
Substituting ${\Gamma}$ in terms of ${\Pi}$ from eq.({\ref{Z}}) in the above equation leads to
\begin{eqnarray}
{\Pi} = -3{\zeta}H\left(1-\frac{\mu n}{\rho+p}\right).
\label{en12}
\end{eqnarray}
Using the Euler relation (\ref{AA}) in the above equation yields 
\begin{eqnarray}
{\Pi} = -3{\zeta}_{e}H
\label{euler}
\end{eqnarray}
where 
\begin{eqnarray}
\zeta_{e} = \left(\frac{nTs}{\rho+p}\right)\zeta~.
\label{eu1}
\end{eqnarray}
From the above relation and eq.({\ref{Z}}), we have
\begin{eqnarray}
\frac{\Gamma}{3H} = -\frac{3\zeta_{e}}{\rho+p}H.
\end{eqnarray}
Using this in eq.({\ref{C}}) and using eq.(s)({\ref{A}}) and ({\ref{E}}) gives
\begin{eqnarray}
\label{EM}
2\dot{H} = -3{\gamma}H^2+3{\zeta}_{e}{H}.
\end{eqnarray}
Now in \cite{Subenoy} it has been shown that for the chemical potential 
${\mu} = 0$ there exists an emergent universe scenario in the first order formalism
of non-equilibrium thermodynamics developed in \cite{Eckart}. This can be seen by putting ${\mu} = 0$ in eq.(\ref{AA}) and then using eq.(\ref{eu1}) to give  ${\zeta}_{e} = {\zeta}$. Eq.(\ref{EM}) then simplifies to
\begin{eqnarray}
2\dot{H} = -3{\gamma}H^2+3{\zeta}{H}~.
\label{gma}
\end{eqnarray} 
It is then easy to see that the solution of this equation gives rise to an emergent universe scenario \cite{Subenoy}. However it is not clear whether there exists an emergent universe scenario for ${\mu}$ ${\neq}$ 0. This is because in this case we have from eq.(\ref{AA}) 
\begin{eqnarray}
\frac{nTs}{\rho +p}=1-\frac{\mu n}{\rho +p}
\label{newe}
\end{eqnarray}
which in turn implies (from eq.(\ref{eu1}))
\begin{eqnarray}
\zeta_{e} = \left(1-\frac{\mu n}{\rho+p}\right)\zeta~.
\label{eu10}
\end{eqnarray}
It is not clear from this expression whether $\zeta_{e}$ is a constant or not. 
We now proceed to show that $\zeta_{e}$ is indeed a constant. 

\noindent The first step is to use eq.(s)(\ref{Z}) and (\ref{E}) to recast the energy conservation eq.(\ref{K}) in the form 
\begin{eqnarray}
\label{37}
{\frac{\dot{\rho}}{\gamma{\rho}}}+3{\frac{\dot{a}}{a}} = {\Gamma}.
\end{eqnarray}
Comparing this equation with eq.({\ref{Y1}}), we get
\begin{eqnarray}
\label{38}
{\frac{\dot{n}}{n}} = {\frac{\dot{\rho}}{\gamma{\rho}}}~.
\end{eqnarray}
Solving this equation yields
\begin{eqnarray}
\label{39}
{\rho}(n) = {K}{n}^{\gamma}
\end{eqnarray}
where $K$ is an integration constant. We also need to find an evolution equation for the temperature in the matter creation scenario.
To do so we first consider ${\rho}$ = $\rho(T,n)$ by regarding $T$ and $n$
as  the basic thermodynamic variables. Using this along with eq.(s)(\ref{Y})
and (\ref{K}) yields
\begin{eqnarray}
\label{270}
\dot{T} = -{\frac{1}{({\partial{\rho}}/{\partial{T})}_n}}[({\rho}+p-{n}{({\partial{\rho}/{\partial{n}})_{T}}){\Phi}+{\Pi}{\Phi}+({\partial{\rho}}/{\partial{n}})_{T}\psi}]
\end{eqnarray}
where $\Phi=\nabla_{\mu}u^{\mu}=3H$ is the fluid expansion \cite{Calvao}.
On the other hand, the fact that $ds$ is an exact differential leads to \cite{Weinberg}
\begin{eqnarray}
\label{280}
{\rho}+p-{n}{({\partial{\rho}}/{\partial{n}})}_{T} = T{({\partial{p}}/{\partial{T}})}_{n}.
\end{eqnarray}
From eq.(s)(\ref{270}) and (\ref{280}), we arrive at the following relation 
\begin{eqnarray}
\label{290}
{\frac{{\dot{T}}}{T}} = -({\partial{p}}/{\partial{\rho}})_{n}{\Phi}-{\frac{{\Pi}{\Phi}+({\partial{\rho}}/{\partial{n}})_{T}\psi}{{T}({\partial{\rho}}/{\partial{T}})_{n}}}~.
\end{eqnarray}
Now using the relation for $\Pi$ in terms of $\psi$ (\ref{Z}) which holds in the isentropic case ($\dot{s} = 0$) and then using eq.(\ref{280}), the above equation takes a simple form
\begin{eqnarray}
\label{30}
{\frac{\dot{T}}{T}} = \left({\frac{\partial{p}}{\partial{\rho}}}\right)_{n}{\frac{\dot{n}}{n}}~.
\end{eqnarray}
Using the barotropic equation of state (\ref{E}) in the above equation, we get 
\begin{eqnarray}
\label{40}
{\frac{\dot{T}}{T}} = {(\gamma-1)}{\frac{\dot{n}}{n}}~.
\end{eqnarray}
Integrating this we get
\begin{eqnarray}
\label{41}
T(n)= {C}{n}^{\gamma-1}
\end{eqnarray}
where $C$ is an integration constant.
Now plugging these expressions for ${\rho}(n)$ and ${T(n)}$ in eq.(\ref{eu1}), we get
\begin{eqnarray}
\label{42}
{\zeta}_{e} = \left({\frac{{C}{s}}{{K}{\gamma}}}\right){\zeta}~.
\end{eqnarray}
This clearly shows that for an isentropic process, ${\zeta}_{e}$ is also a constant. Hence eq.(\ref{EM}) once again has an emergent universe solution. This can be seen by solving eq.(\ref{EM}) to yield
\begin{eqnarray}
H(t) = \frac{\zeta_e e^{\frac{3\zeta_e}{2}(t+C_1)}}{1+\gamma e^{\frac{3\zeta_e}{2}(t+C_1)}}
\end{eqnarray}
where $C_1$ is a constant of integration.
Integrating once more we get
\begin{eqnarray}
a(t) = C_2 \left(1+\gamma  e^{\frac{3\zeta_e}{2}(t+C_1)}\right)^{\frac{2}{3\gamma}}
\end{eqnarray}
where $C_2$ is a constant of integration. This expression for $a(t)$ has the properties
of an emergent universe \cite{SGPaper20}.
Therefore the first order theory \cite{Eckart} allows an emergent universe scenario even when the chemical potential is non-vanishing.

%




\section{Second order formalism}

In this section we shall discuss that there exists no emergent universe scenario
in the second order formalism of non-equilibrium 
thermodynamics \cite{Israel, Stewart}. 

\noindent The entropy flow vector in the second order theory (considering deviations away from the equilibrium) is chosen to be \cite{Israel, Stewart} 
\begin{eqnarray}
s^\mu = \left[sn-\frac{\tau\Pi^2}{2\zeta{T}}\right]u^\mu
\label{so1}
\end{eqnarray}
where $\tau$ is the relaxation time that determines the relaxation from states with $\dot{s}>0 $ to those with $\dot{s} = 0$. It is to be noted that this theory is a causal theory in contrast to the first order theory (discussed earlier) which is a non-causal theory. The introduction of a finite relaxation time makes this theory physically different from the first order theory.

\noindent The divergence of the entropy reads \cite{Israel, Stewart}
\begin{eqnarray}
\nabla_{\mu}{s^\mu} = -\frac{\Pi}{T}\left[3H+\frac{\tau}{\zeta}\dot{\Pi}+\frac{1}{2}\Pi{T}\nabla_{\mu}\left(\frac{\tau}{\zeta{T}}u^\mu\right)\right].
\label{so2}
\end{eqnarray}
Once again for the second law of thermodynamics to hold, one chooses
\begin{eqnarray}
\Pi = -\zeta\left[3H+\frac{\tau}{\zeta}\dot{\Pi}+\frac{1}{2}\Pi{T}
\nabla_{\mu}\left(\frac{\tau}{\zeta{T}}u^\mu\right)\right].
\label{so3}
\end{eqnarray}
This therefore gives $\nabla_{\mu}{s^\mu}=\Pi^2/(\zeta T)\geq 0$. 
As a consequence of this the effective bulk viscous pressure $\Pi$ now 
becomes a dynamical variable and has the following causal evolution equation
\cite{Israel, Stewart}
\begin{eqnarray}
\Pi +\tau \dot{\Pi}= -3\zeta H-\frac{1}{2}\Pi \tau \left[3H +\frac{\dot{\tau}}{\tau}-\frac{\dot{\zeta}}{\zeta}-\frac{\dot{T}}{T}\right]~.
\label{so4}
\end{eqnarray}
As $\tau\rightarrow 0$, $\Pi$ does not remain a dynamical variable any more and  reduces to the usual relation $\Pi = -3\zeta{H}$. It is also a simple matter to see that the evolution equation for $\Pi$ is difficult to solve due to its complicated nature. However, it was argued in \cite{Subenoy} that there still
exists an emergent universe scenario in this formalism if one chooses the matter creation rate $\Gamma$ to be a constant phenomenologically 
and then solve eq.(\ref{C}) which leads to
an emergent universe solution. 

\noindent In this paper we would like to mention that there is a problem with this argument which can be seen as follows. It should be noted that eq.(\ref{C})
follows from the isentropic condition (\ref{Z}). This implies that choosing
the matter creation rate $\Gamma$ to be a constant phenomenologically fixes $\Pi = -3\zeta{H}$ which follows from eq.(\ref{Z}). This in turn implies taking the $\tau\rightarrow 0$
limit which no longer renders $\Pi$ to be a dynamical variable.
Therefore, it is evident that there exists no emergent scenario in the second order formalism of non-equilibrium thermodynamics.



\section{Duality between matter creation and equation of state for emergent universe}
In this section we discuss a duality between matter creation with barotropic equation of state and the non-linear equation of state which gives rise to an emergent universe scenario. For the non-linear equation of state \cite{SGPaper20}
(which gives rise to an emergent universe scenario)
\begin{eqnarray}
p=A\rho-B\sqrt{\rho}
\label{nonl}
\end{eqnarray}
where $A$ and $B$ are constants, one gets from eq.(s)(\ref{A}) and (\ref{B})
\begin{eqnarray}
2\dot{H} = {-(3A-1)H^2}+\sqrt{3}BH.
\label{du1}
\end{eqnarray}
Comparing it with eq.(\ref{EM}) (which also leads to an emergent universe solution) obtained in a matter creation scenario with a barotropic equation of state ({\ref{E}}), one gets
\begin{eqnarray}
A = \gamma+\frac{1}{3}
\label{du2}
\end{eqnarray}
\begin{eqnarray}
B = {\sqrt{3}}{\zeta}_{e}~.
\label{du3}
\end{eqnarray}
Thus we find that a non-linear equation of state (\ref{nonl}) with the constants $A$ and $B$ being identified with the adiabatic index $\gamma$ and the bulk viscosity coefficient ${\zeta}_{e}$ is identical to a universe with matter creation  and a barotropic equation of state (\ref{E}).

\section{Conclusion}
In this paper we have obtained an emergent universe scenario in the most general case of the chemical potential being non-zero in the first order non-equilibrium thermodynamics with matter creation. It is worth mentioning that such a solution was shown to exist under a restricted condition, namely, for vanishing chemical potential. We have then argued that there exists no emergent universe solution in the second order theory of non-equilibrium thermodynamics which is a causal theory. Finally we demonstrate a duality between matter creation for barotropic fluid and equation of state for emergent universe. 

\section*{Acknowledgements}
S.G. acknowledges the support by DST SERB under Start Up Research Grant (Young Scientist), File No.YSS/2014/000180.

\end{document}